# Influence of an adsorbing polymer in the aging dynamics of Laponite clay suspensions

L. ZULIAN[†§*], B. RUZICKA[†§] and G. RUOCCO[†§]

† Dipartimento di Fisica, Università di Roma "La Sapienza", P.le Aldo Moro 5, I-00185 Roma (Italy)
§ INFM-CRS SOFT, Università di Roma "La Sapienza", P.le Aldo Moro 5, I-00185 Roma (Italy)

*Corresponding author. Email: Laura.Zulian@roma1.infn.it

## ABSTRACT

Clay-polymer dispersions in aqueous solutions have attracted a great interest in recent years due to their industrial applications and intriguing physical properties. Aqueous solutions of bare Laponite particles are known to age spontaneously from an ergodic state to a non ergodic state in a time varying from hours to months depending on Laponite concentration. When a polymer species like Polyethylene Oxide (PEO) is added to the solution, it weakly adsorbs on clay particle surfaces modifying the effective interaction potential between Laponite particles. A dynamic light scattering study, varying polymer concentration at fixed polymer molecular weight ($M_w$=200.000 g/mol), has been performed in order to understand the effect of polymer on the aging dynamics of the system. The results obtained show that arresting phenomena between clay particles are hindered if PEO is added and consequently the aging dynamics slows down with increasing PEO concentration. This process is possibly due to the progressive coverage of the clay surface by polymers that grow with increasing PEO concentration and may lead to steric stabilization.

# § 1. Introduction

Clay-polymers complexes have technological interest since they find application in a wide range of industrial formulations as rheological modifiers or stabilizers. Moreover in the scientific community the interaction between clay and polymer has received a considerable attention especially due to the fact that polymers can be used as fine-tuners of the interaction potential between colloidal particles giving the possibility to have a deeper comprehension of the mechanism that originates structural arrested states like gel and glass [1], [2] Laponite is a synthetic hectorite clay where single particles have a characteristic discoidal shape with diameter of 25 nm and thickness of 1 nm [3]. In aqueous dispersions Laponite clay particles show a strongly negative charge on the face and a weakly positive/negative charge on the rim. The competition between attractive and repulsive interactions causes the aqueous dispersions of Laponite to have a non-trivial behavior also at very low clay concentrations [4-11]. In fact clay dispersions spontaneously *age* from an initial liquid like state towards a final arrested solid-like structure that is gel-like or glass like depending on Laponite concentration [7]. The fundamental physics leading to interaction between clay particles, the mechanism of gel/glass formation and the role of the attractive and repulsive interactions remain a subject of debate and need to be better understood [4-11].

A simple method to clarify the role of attractive and repulsive terms in the interaction potential of Laponite systems is to modify the relative contribution of the two interaction terms and to study its effect on aging phenomenon and on the realization of two final arrested state. The addition of a water soluble adsorbing polymer like Polyethylene Oxide (PEO) is expected to affect kinetic behavior of Laponite dispersions [12-16]. A Small Angle Neutron Scattering study of this kind of system [17] has shown that PEO adsorbs around Laponite particles. Fixing molecular weight and increasing PEO concentration, the effective coverage surface of clay discs increases, bringing the system towards a steric stabilization of the colloidal dispersion. Also steric hindrance due to excluded volume between polymers acts against the formation of an arrested state. Therefore increasing PEO concentration the arrest process becomes more and more efficiently hindered and in the limit case could be completely avoided.

A dynamic light scattering study on the arresting kinetics of Laponite/PEO mixtures at fixed clay and polymer concentrations varying molar mass of Polyethylene Oxide has been performed by Nelson et al. [13]. The authors find that the addition of polymers decreases the rate of aggregation due to the addition of polymer due to steric stabilization coming from polymer adsorption. Nevertheless the authors do not quantify the effect in relation to different amounts of added polymer and no systematic studies are present in literature. Finally

Nelson and coworkers add salt to the solution which causes a further modification in the interaction potential, decreasing the characteristic screening length of the repulsive part of the potential.

In this paper we report a dynamic light scattering study about the influence of addition of PEO at fixed morecular weight ($M_w$=200.000 g/mol) on the kinetics of an arrested state formation of Laponite system.

We observe a slowing down of the characteristic aging dynamics of Laponite systems that clearly depends on PEO concentration. However the formation of a final arrested state is never avoided, even at the highest polymer concentration investigated.

§ 2. Experimental details

Laponite RD is produced by Laporte Industries with a density of 2570 Kg/m$^3$ and used as received. Each crystal is composed of roughly 1500 unit cells with an empirical chemical formula of $Na^+_{0.7}[(Si_8Mg_{5.5}Li_{0.3})O_{20}(OH)_4]^{-0.7}$. The clay powder is composed of single crystals with a three layers structure: the central one is constituted by magnesium and litium atoms bonded with oxigen and OH groups to form an octahedral structure. This layer is sandwiched within two tethraedral structures of silicon with sodium atoms that are exposed on the surface of the platelets [3]. When this kind of chemical architecture is dispersed in water, it releases Na ions from the surface because of ionization phenomenon, originating a net negative charge of roughly some hundreds of elementary charges. On the contrary, due to a protonation phenomenon of the OH groups disposed on the rim of the plates, a weakly positive charge appears. Therefore in water Laponite clay forms a colloidal dispersion of charged disc like particles with a diameter of 25 nm and a thickness of 1 nm that interact with both attractive and repulsive interactions. Polyethylene oxide (PEO) with a molecular weight of $M_W$=200.000 g/mol is supplied from Sigma Aldrich and used as received. The general chemical formula of PEO is -[$CH_2CH_2O$]-$_n$. It is a linear chain obtained by a polymerization of ethylene oxide. Using the empirical relation $\ell \approx M_w^{0.57} / 18.4$ reported in [18], [19] we can estimate the characteristic linear length of the chosen polymer to be $\ell \approx$ 57 nm, comparable with the characteristic dimensions of a Laponite disc.

PEO stock solution at $C_{PEO}$ =1 % was prepared adding a weighted quantity of polymer in 18 MΩ high pure deionized water and stirring the dispersion at room temperature for several hours. Afterwards the stock solution was diluted adding deionized water in order to obtain PEO solution at two different weight percentages ($C_{PEO}$=0.4% and $C_{PEO}$=0.5%). Laponite/PEO samples were prepared by fully dispersing in this solution the desired percentage of Laponite powder at fixed concentration $C_w$=2.0% in weight. Laponite/PEO dispersions were then stirred vigorously for 30 minutes. The starting aging (or waiting) time ($t_w = 0$) is defined as the time when the stirring process is ended.

Dynamic light scattering (DLS) measurements were performed using an ALV-5000 logarithmic correlator in combination with a standard optical setup based on an He-Ne ($\lambda$=632,8 nm) 10 mW laser. The intensity correlation function was directly obtained as:

$$g_2(q,t) = \frac{<I(q,t)I(q,0)>}{<I(q,0)>^2} \tag{2.1}$$

where q is the modulus of the scattering wave vector defined as $q = 4\pi\, n\, /\lambda\, sin(\theta/2)$ with $\theta=90°$ in this experiment. The dynamic structure factor f(q,t) can be directly obtained inverting the Siegert relation :

$$f(q,t) = \sqrt{\frac{g_2(q,t) - 1}{b}} \tag{2.2}$$

where *b* represents the coherence factor. All measurement have been performed at room temperature.

## § 3. Results and discussion

Dynamic light scattering measurements at fixed Laponite concentration ($C_W$=2.0 %) for two different PEO concentrations ($C_{PEO}$=0.4 % and $C_{PEO}$=0.5 %) with fixed molecular weight $M_W$=200.000 g/mol added to the dispersion have been performed at several waiting times in order to clarify the effect of the polymer on the aging dynamics. As an example, correlation functions at increasing waiting times $t_w$ for neat Laponite sample and for the sample at $C_W$=2.0% and $C_{PEO}$=0.4% are shown in Fig. 1.

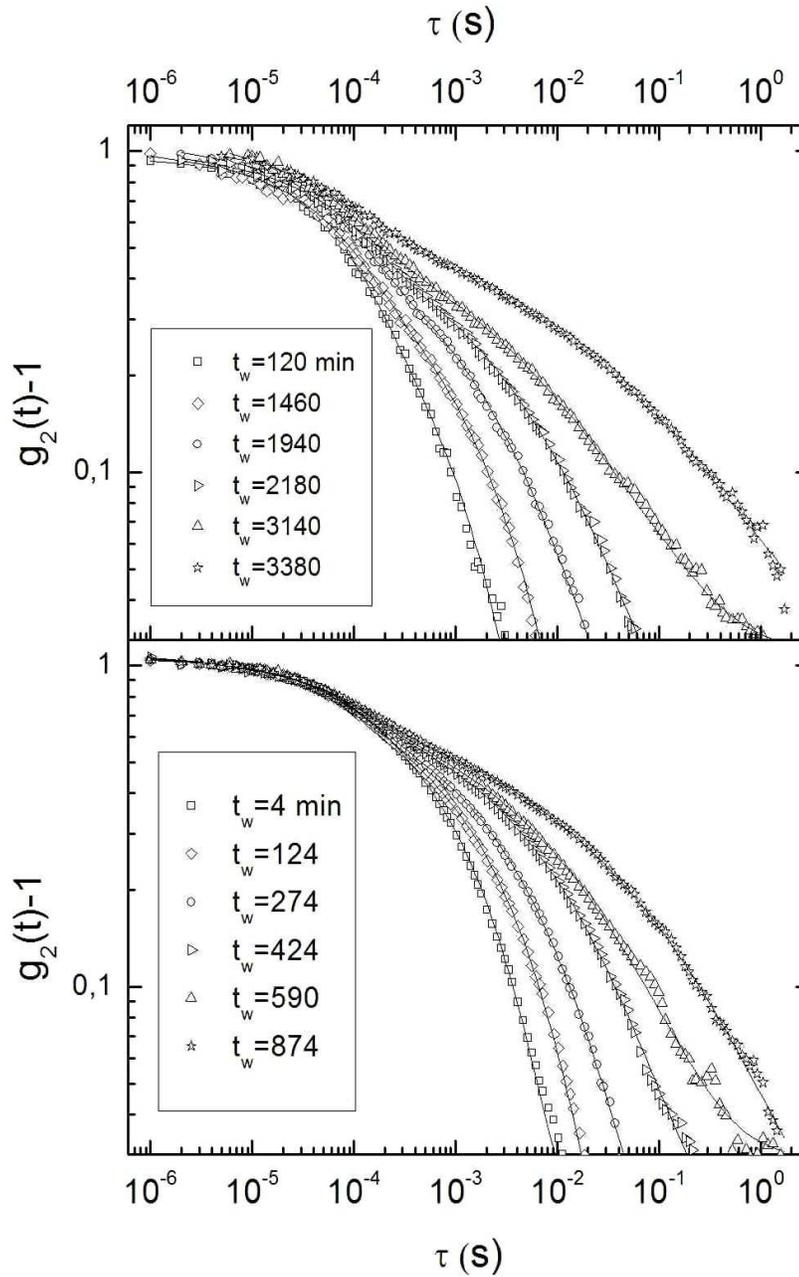

***Figure 1:***
*Waiting time evolution of autocorrelation functions for neat Laponite sample at $C_w = 2.0\%$ (top panel) and for $C_w = 2.0\%$, $C_{PEO} = 0.4\%$ Laponite-PEO dispersion sample (bottom panel). Solid lines represent fits obtained through Eq. 3.1.*

It is evident from the figure that both systems undergo *aging*, i.e. the characteristic dynamical properties of the sample change with waiting time reflecting the structural rearrangement inside the sample during the transition from an ergodic to an arrested non-ergodic state. In particular dynamics becomes slower and slower for increasing waiting time $t_w$.

The correlation functions (Fig. 1) decay following a two steps behavior: two different relaxation processes, a fast and a slow ones, are present in the system on two different time scales in all the investigated samples, with and without added polymer. Therefore adding a polymer species to Laponite dispersions do not modify drastically the characteristic shape of autocorrelation functions. For this reason we can use the same quantitative analysis we have already used in the past for neat Laponite samples [5], [6] and we fit the $g_2(t)-1$ behavior using a sum of simple and stretched exponential functions [20]:

$$g_2(q,t)-1 = b(a \cdot e^{-\frac{t}{\tau_1}} + (1-a)e^{-\left(\frac{t}{\tau_2}\right)^\beta})^2 \qquad (3.1)$$

Where $b$ represents the coherence factor and $a$, $\tau_1$, $\tau_2$ and $\beta$ are shape fitting parameters.

As an example the dynamic structure factor f(q,t) (open circles) for the sample at $C_w$ =2.0 % and $C_{PEO}$=0.4 % at $t_w = 3.00 \cdot 10^4$ s, the fit (full line) and the two contributions (dotted and dashed lines) are shown in Fig. 2. From this Figure the goodness of the fit is evident moreover one can easily identify how the two different terms of Eq. 3.1 contribute to the fit. The first decay, that describes the fast dynamics ($\tau_1$) on short time scale is related to a vibrational diffusive mode of the single particle inside the cage of its neighbors and that is well described by a single exponential curve (dotted line of Fig. 2). Slow dynamics on longer time scales, that is manifestly non-Debye, is instead described by a stretched exponential function (with $\tau_2$ and $\beta$ parameters) and is shown in Fig. 2 as dashed curve. This second decay is related to structural rearrangements and to the strong complexity of the system. The fits are shown in Fig. 1 as solid lines.

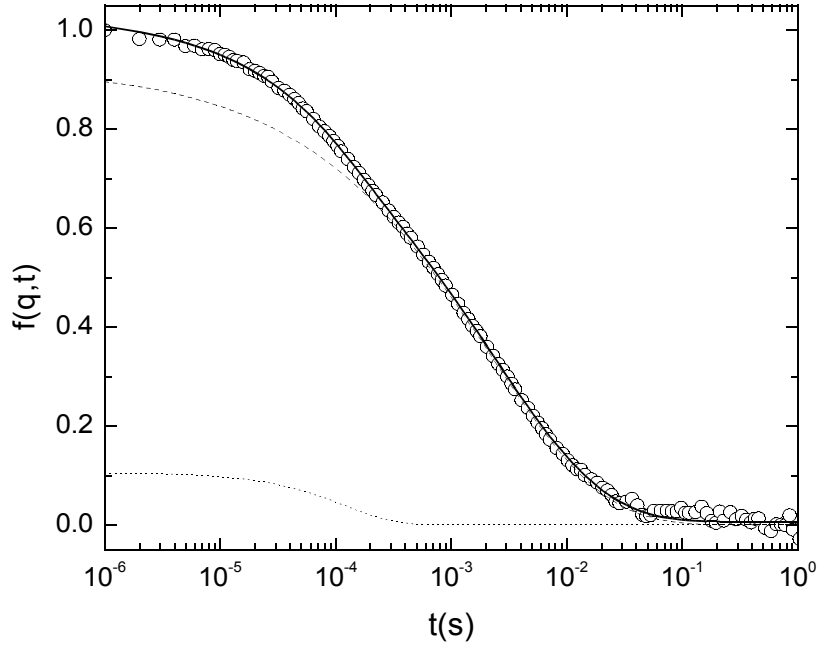

***Figure 2:*** *Dynamic structure factor f(q,t) (Eq. 2.2) (open circles) for the sample at $C_w$= 2.0% and $C_{PEO}$=0.4% at waiting time $t_w$= 3.00 · $10^4$ s and corresponding fit with Eq. 3.1 (full line) with simple exponential (dotted line) and stretched exponential (dashed line) contributions.*

The dependence of the parameters obtained from the fit of time correlation functions on waiting time $t_w$ permit to obtain quantitative informations about the influence of added PEO on Laponite dispersions. To focus the attention on the slow dynamics parameters we can define an useful quantity that takes into account both significative parameters of slow dynamics: relaxation time ( $\tau_2$ ) and stretching parameter ( $\beta$ ). We define then the *mean relaxation time* [5]:

$$\tau_m = \frac{\tau_2}{\beta}\Gamma\left(\frac{1}{\beta}\right) \tag{3.2}$$

In Fig. 3 the waiting time behavior of $\tau_m$ for samples at fixed Laponite concentration ($C_w$=2.0 %) and for different PEO concentrations ($C_{PEO}$=0 % (neat Laponite sample) filled circles; $C_{PEO}$=0.4 %, open squares; $C_{PEO}$=0.5 %, open triangles) are shown.

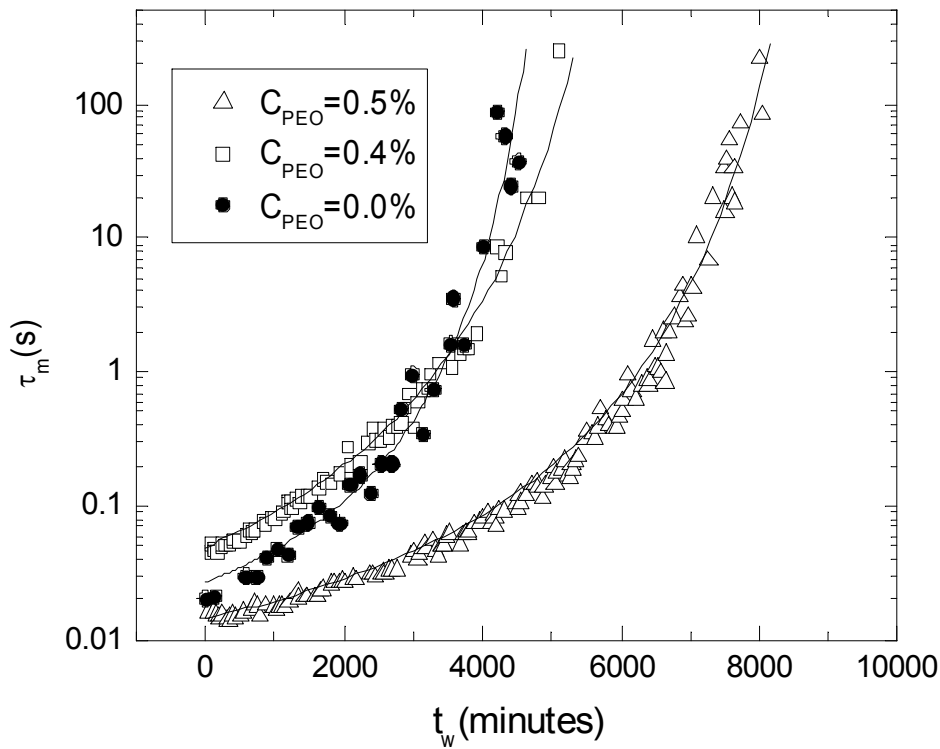

*Figure 3:* *Waiting time evolution of $\tau_m$ for samples with fixed Laponite concentration ($C_w$= 2.0%) without added PEO (full circles) and for two different PEO concentrations ($C_{PEO}$=0.4%, open squares; $C_{PEO}$=0.5%, open triangles). More PEO is added to the dispersion more the dynamics of the system toward an arrested state is slowed down. Solid lines are fits obtained through Eq. 3.3.*

As already observed in our previous studies on neat Laponite dispersions all samples show a characteristic exponential growth of the mean relaxation time $\tau_m$ which reflects the progressive slowing down of the dynamics. At a certain waiting time, that we call $t_w^\infty$, a divergence appears [5]. At this time the raw autocorrelation spectra have undergone a crossover from a complete to an incomplete decay. This behavior is commonly identified as the indication of an ergodicity breaking, signature of a transition towards an arrested state [5],[21],[22]. Therefore this critical waiting time value can be considered as a good estimation (in excess) of the time the system needs to undergo a complete transition from a liquid like ergodic state towards a solid like non ergodic arrested state. It has been observed that $t_w^\infty$ depends on Laponite concentration, i.e. higher is clay concentration less time is spent by the system to age and to complete the whole ergodic to non ergodic transition [5], [8]. Moreover at fixed Laponite concentration, the addition of salt in Laponite dispersions causes an abrupt decrease in the characteristic value at which the divergence in the $\tau_m$ behavior appears [6]. In the following we will study what happens to this value when an increasing quantity of PEO is added. We can obtain the $t_w^\infty$ parameter for each investigated sample by fitting the waiting time behavior of mean relaxation time whit a phenomenological law [5]:

$$\tau_m = \tau_0 \exp\left[\frac{B}{1-\frac{t_w}{t_w^\infty}}\right] \qquad (3.3)$$

where B, $\tau_0$ and $t_w^\infty$ are fit parameters. The fitting curve obtained trough Eq.3.3 are shown in Fig. 3 as solid lines and well describe also the samples with added PEO. In Fig. 4 the behavior of $t_w^\infty$ in function of PEO

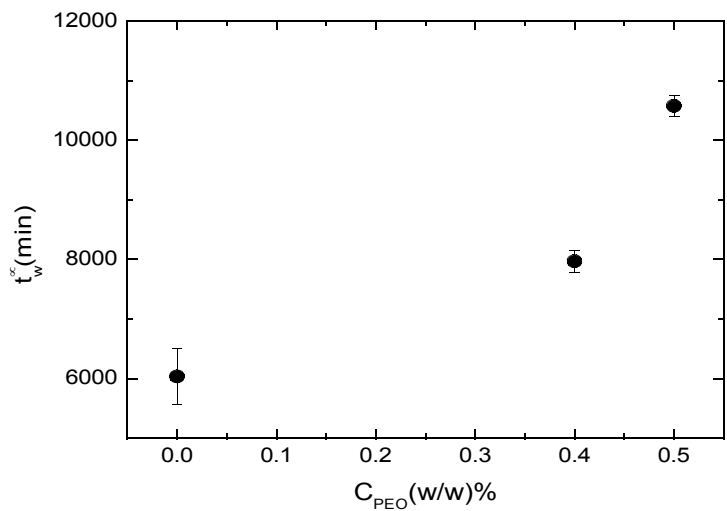

*Figure 4:* $t_w^\infty$ *parameter for samples with Laponite concentration $C_w$= 2.0% at different PEO concentrations. The increasing of PEO concentration cause an increase in this characteristic critical waiting time: the arrest becomes more and more hindered and therefore the gelation kinetics slow down.*

concentration shows an increase in the $t_w^\infty$ parameter as PEO amount is increased.

More in details samples with $C_{PEO}$=0.4 % and $C_{PEO}$=0.5 % need respectively ~1.3 and ~1.8 times the time needed by neat Laponite sample (at the same clay concentration) to complete age towards the final arrested state. This is due to the presence of the adsorbing polymer: more PEO is added to the dispersion more efficient becomes the coverage of clay surface and then the barrier against aggregation. Consequently the kinetic towards the arrested state slows down. Nevertheless we do not observe a complete suppression of the aggregation at the studied PEO concentrations. Moreover due to the characteristic length of the polymer used ($M_W$=200.000 g/mol, $\ell\approx$57 nm) comparable with the interparticle distance between two different Laponite particles, the bridging phenomenon cannot be ruled out and could be present in the system. With our study we cannot estimate the exact contribution of the bridging phenomenon to the slowing down of the dynamics but we can obtain some useful information from the fast relaxation time $\tau_1$.

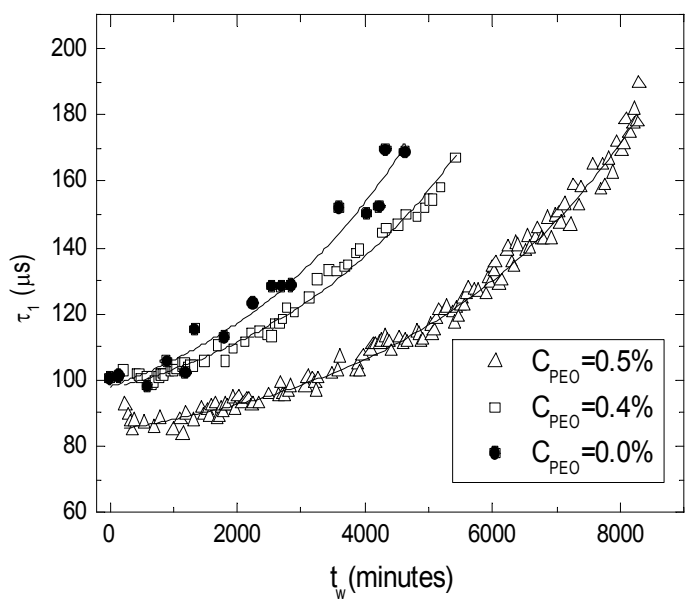

*Figure 5:* Waiting time evolution of $\tau_1$ for samples with fixed Laponite concentration ($C_w$= 2.0%) for neat Laponite sample (full circles) and for two different PEO concentrations ($C_{PEO}$=0.4% , open squares; $C_{PEO}$=0.5% open triangles). The solid lines represents fits with a simple exponential function.

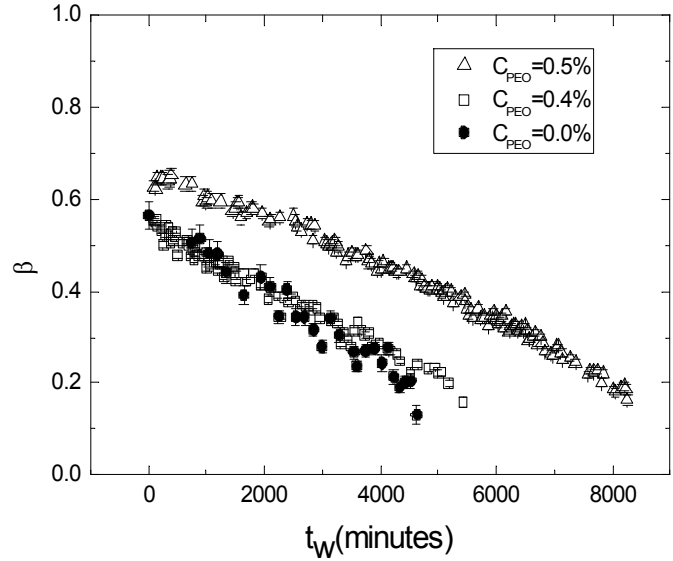

***Figure 6:*** *Waiting time evolution of the stretching parameter β for samples with fixed Laponite concentration ($C_w$ = 2.0%) for neat Laponite sample (full circles) and for two different PEO concentrations ($C_{PEO}$=0.4% , open squares; $C_{PEO}$=0.5 % open triangles).*

This time can be related to the vibrational motion and the characteristic dimensions of single particle inside the cage of neighbors. The behavior of $\tau_1$ is shown in Fig. 5 and its increase at increasing waiting time for all samples investigated is evident. Fitting this behavior with a single exponential function we can obtain informations about the value of the $\tau_1$ parameter at the starting waiting time $t_w$=0, that we call $\tau^0_1$. This value is an indication of the dimensions of the characteristic objects inside the dispersion. We also define a dimensionless time constant $\tau^* = \tau^0_1/\tau_0$; where $\tau \approx 76$ μs, is the decay time measured for the bare Laponite particles in solution at $t_w$=0. The values of $\tau^0_1$ found for samples with $C_{PEO}$=0.4 % and $C_{PEO}$=0.5 % are $\tau^0_1$=79 μs and $\tau^0_1$=74 μs respectively. Then the dimensionless quantity $\tau^* \approx 1$. This indicates that the dynamics are indistinguishable from that of bare Laponite in solution. Since the polymer adsorbs on the face of the discs rather than on the edges [13], there is no increase in the longest dimension of the discs, and there is no change in the translational diffusion coefficient, so that the decay time for Laponite particles with adsorbed polymer is the same as for the bare Laponite particles, and $\tau^* \approx 1$. These arguments suggest that at least for this polymer concentration, the formation of bridging between particles is not probable.

Looking at Figure 5 there is also a clear dependence of $\tau_1$ from polymer concentration, in agreement with what observed for the mean relaxation time $\tau_m$ (see Fig. 3).

In Figure 6 the waiting time behavior of the stretching parameter $\beta$ is shown for all samples investigated ($C_{PEO}$=0 % (neat Laponite sample) filled circles; $C_{PEO}$=0.4 % , open squares; $C_{PEO}$=0.5 %, open triangles). For all samples the value of the stretching parameter decreases with time from a value of ~0.7 to an extremeli low value of ~0.2. This quantifies the behavior of the raw photoncorrelation spectra shown in Fig. 1 where it is evident that at increasing waiting times the curves become more and more stretched. Also in this case the dependence on polymer concentration is clear and consistent with the behavior found for the mean relaxation time $\tau_m$ and for the fast relaxation time $\tau_1$.

## § 4. Conclusions

Dynamic light scattering measurements have been performed on aqueous Laponite samples varying Polyethylene Oxide concentration added to the dispersion (at a fixed molecular weight of $M_w$=200.000 g/mol). In conclusion we have shown that addition of PEO modifies the gelation kinetics respect to pure Laponite sample. Nevertheless the characteristic shape of raw autocorrelation functions do not show drastic changes, then we can apply the same fit analysis we introduced and used previously for investigation on Laponite clay in water [5],[7] and in water with the addition of NaCl salt [6]. The binding between different Laponite particles is inhibited by the addition of this polymeric species that is known to adsorb on clay surface. This leads to a slowing down in the formation of Laponite final arrested state. We have also shown a clear dependence of this phenomenon on PEO concentration. This is in agreement whit the idea that increasing PEO concentration corresponds to a growth in the coverage surface of Laponite discs by adsorbed polymer and then to an increase in the repulsive term due to steric hindrance that hinder more and more the dynamical arrest. An extensive study in order to understand more in details the dependence of aging on PEO and molar mass of PEO is ongoing.